\documentclass[12pt]{article}
\usepackage[a4paper, twoside, hmarginratio=1:1, vmarginratio=1:1, left=1in,top=1in]{geometry}
\usepackage{amsfonts}
\usepackage{graphicx}
\usepackage{epstopdf}
\usepackage{algorithmic}
\usepackage{amsmath}
\usepackage{amssymb}
\usepackage{amscd}
\usepackage{epsfig}
\usepackage{graphics}
\usepackage{enumitem}
\usepackage{natbib}
\usepackage{float}
\usepackage{bm}
\usepackage{multirow}
\usepackage{verbatim}
\usepackage{lineno,hyperref}
\usepackage{algorithm}  
\usepackage{algorithmic}

\begin{document}

\title{Efficient Social Distancing for COVID-19: An Integration of Economic Health and Public Health}

	\author{Kexin Chen\thanks{Department of Statistics, The Chinese University of Hong Kong, Shatin, N.T., Hong Kong. (\tt{kxchen@link.cuhk.edu.hk}).}
	\and Chi Seng Pun\thanks{School of Physical and Mathematical Sciences, Nanyang Technological University, Singapore.(\tt{cspun@ntu.edu.sg}).}
	\and Hoi Ying Wong\thanks{Department of Statistics, The Chinese University of Hong Kong, Shatin, N.T., Hong Kong. (\tt{hywong@cuhk.edu.hk}).}}
\maketitle                  





\begin{abstract}
Social distancing has been the only effective way to contain the spread of an infectious disease prior to the availability of the pharmaceutical treatment. It can lower the infection rate of the disease at the economic cost. A pandemic crisis like COVID-19, however, has posed a dilemma to the policymakers since a long-term restrictive social distancing or even lockdown will keep economic cost rising. This paper investigates an efficient social distancing policy to manage the integrated risk from economic health and public health issues for COVID-19 using a stochastic epidemic modeling with mobility controls. The social distancing is to restrict the community mobility, which was recently accessible with big data analytics. This paper takes advantage of the community mobility data to model the COVID-19 processes and infer the COVID-19 driven economic values from major market index price, which allow us to formulate the search of the efficient social distancing policy as a stochastic control problem. We propose to solve the problem with a deep-learning approach. By applying our framework to the US data, we empirically examine the efficiency of the US social distancing policy and offer recommendations generated from the algorithm.\\

\end{abstract}

\newpage


\section{Introduction}

The coronavirus disease (COVID-19), which was first identified in Wuhan, China, in Dec 2019, has resulted in a pandemic. Its widespread outbreak since Feb 2020 imposes unexpected challenges on countries around the globe, as it has caused hundreds of thousands of deaths, severe resource shortages, and a severe economic blow. Several studies estimate an inevitable global recession as a result of the pandemic. \cite{WrenLewis2020} suggests a decline in economic growth due to disrupted global supply chains and reduced social consumption. According to \cite{Fernandes2020}, the GDP growth could decline by 3-5\% depending on the country, costing 2.5-3\% of global GDP for each additional shutdown month. \cite{Makridis2020} estimates a 5\% decline in GDP growth monthly in United States (US), while the economic cost of the first two months since the outbreak of COVID-19 in US has already been \$2.14 trillion, roughly 10\% percent of GDP in US.

A broad variety of mathematical models studying the evolution of (COVID-19) epidemic processes have been proposed in the literature. A classic deterministic compartmental modeling framework, SIR model, is first proposed
in \cite{Kermack1927}. 
Recently, SIR model variations have been developed with increasing mathematical complexity, enabling the study of several aspects of the epidemic process. For example, \cite{Karako2020,Bardina2020} models an additional individual action, including latent periods and the presence of asymptomatic patients. Several studies (\cite{He2020,Chang2020,Moore2020}) model the transmission of COVID-19 by considering government's control measures, while some studies (\cite{Souza2020}) incorporate the epidemiological and clinical characteristics in modeling. The aforementioned studies improve our understanding of the underlying virus transition mechanism and provide guidance on how to manage the spread of the virus. 


Among policies in reaction to pandemic, the non-pharmaceutical social distancing is a crucial tool for minimizing the spread of virus by preserving physical distance and decreasing social interactions; see \cite{Habersaat2020} for ten qualitative considerations in the new norm. Common social distancing laws include bans on public events, close-downs of schools and non-important workplaces, limits on public transports;see \cite{Kissler2020}. Companies like Google, Apple, and Cuebiq have already utilized the location data of their users to deliver useful mobility and economic trends information; see \cite{Google,Apple,Cuebiq}. A vast amount of recent COVID-19 studies focus on researching the pattern of changes in mobility (\cite{Warren2020,Ghader2020,Gao2020}), the role of mobility in disease prediction (\cite{Liu2020,Miller2020}), estimating the effectiveness of distancing social strategies (\cite{Liu2020,Soucy2020}) and economic consequence of mobility restriction (\cite{Bonaccorsi2020}).

While social distancing can effectively lower infection rates, policymakers face a complicated situation in trying to reconcile stringent public health acts and keep the economy running. The combined impact of shutdown policies with public health crisis further results in a sharp contraction of economic activity; see \cite{Barua2020}. Substantial evidence suggests that government restrictions on commercial activities bring huge economic damage, especially for a service-oriented economy like US; see \cite{Baker2020}. \cite{McKee2020} adds that if the world fails to protect the economy, the COVID-19 could cost beyond the disease itself economically and in turn the healthcare system's blow will lead to long-lasting damage in health in the future. An efficient social distancing policy (ESDP) that takes into account both public health and economic health is desired and draws researchers' attention. Some highly related studies are \cite{Benzell2020,Block2020}. However, the aforementioned studies on social distancing are more of qualitative discussion, offering guidance on how to implement social distancing. \cite{Lopez2020} conducts data analysis with real COVID-19 data and conduct simulation studies to evaluate different scenarios. \cite{Niko2020} develop a hybrid forecasting method for infectious disease and the excess demand for products and services during the pandemic. \cite{Silal2020} offers a comprehensive overview of operational research methods for the management of infectious disease. 

However, we notice that the existing studies did not tease out the mobility changes from the stochastic modeling of the COVID-19 processes, given that the data we used to feed into the model have been affected by the time-varying social distancing policies. Moreover, the literature is lack of a quantitative study on the ESDP. The key to these two research gaps is to take advantage of the community mobility data from Google. This paper contributes to the literature in the following aspects. 
\begin{enumerate}
 \item We extend the susceptible, infectious, recovered, and deceased (SIRD) model
 to incorporate community mobility data to predict the COVID-19 process,
while the randomnesses are studied statistically. This enables us to regard the mobility as the control variable.
\item We statistically learn the COVID-driven economic value from a major financial index, the mobility changes, and the COVID-19 statistics.
\item  We then formulate a stochastic control problem to determine the optimal
 mobility target (or social distancing policy) and solve it numerically via a deep learning approach. The objective functional of the stochastic control problem integrates public health and economic health while the state processes are the statistically estimated SIRD model and the COVID-driven economic value. By varying the weights between the public and economic health risk in the objective functional, we define the notion of the efficient social distancing
frontier(ESDF), through which we measure the efficiency of realized social distancing schemes.
\end{enumerate}
Our proposed framework consists of three novel components learnt from the COVID-19 pandemic experience. The first one is the incorporation of social mobility into to the stochastic epidemic model detailed in Section \ref{sec:stomodel}. We empirically show the significance of mobility controls during the COVID-19 crisis period. The second novel idea is to learn the economic cost from the frequently updated financial market data in Section \ref{sec:econ}. We link the stability of major financial indices to the realized epidemic and mobility control variables. Our empirical tests confirm the significance and relevancy of the cost functional to the stochastic epidemic model. The learning of the cost functional is performed for the data before any economic stimulus because our goal is to reduce the economic cost from the disease rather than predicting the index movement for profit making. 
Section \ref{Section4} details the conceptual framework of integrating public health and economic health to formulate ESDPs in \eqref{concept}, results are provided in Section \ref{sec:result}. Finally, Section \ref{sec:con} concludes.


\section{Stochastic modeling of COVID-19} \label{sec:stomodel}
We model the COVID-19 process based on a compartmental epidemiological framework, originated in \cite{Kermack1927}. Specifically, the population is assigned to compartments with labels, $S$, $I$, $R$, and $D$, representing the fractions of individuals who are \textit{susceptible}, \textit{infectious}, \textit{recovered}, and \textit{deceased}, respectively, to the infectious disease. There are also many extensions with more or modified compartments;
see \cite{Hethcote2000}. However, their applicability highly depends on the available data and for most infectious disease, especially COVID-19, the access of the data is limited and the accuracy of some data
is arguable when enormous data are being processed everyday. Therefore, we only focus on the SIRD model that only require the number of reported cases, active cases, and deaths.

The stochastic modeling in this paper consists of two parts. In the first part, we aim to build a stochastic SIRD model that can fit with the real data of COVID-19 in United States (US). Although the SIRD model is usually presented in differential equations, we consider a discrete-time SIRD model as it is more intuitive in practice. To align with the frequency of the data we collect, the time interval is set as one day, indexed by $t\in \mathbb{N}$. A standard
 SIRD epidemiological model is also used in \cite{fernandez2020estimating} to inversely back out a time-varying virus reproduction rate to capture changes in behavior and policy that occur at different times. However, we would like to use real data to estimate the stochastic process of the reproduction rate.  This research demonstrates the value of modeling the changing infection rates of social distancing and enhanced hygienic practices. The distinct feature of our modeling of COVID-19 is to incorporate with the Google mobility indices, available at \cite{Google}, into the model and our studies show that they are highly correlated with the infection rates. In fact, without taking the mobility of the community into consideration, the modeling of the infectious disease is inaccurate. In the next section, we further build an econometric model on the market indices with the independent variables using COVID-19 statistics and the Google mobility indices. We also find that the COVID-19 and mobility indices play important roles in explaining the market indices during the COVID-19 crisis.

\subsection{The SIRD model with stochastic log-odds} \label{sec:model}

Now, the $S$, $I$, $R$, and $D$ are treated as (stochastic) processes indexed by $t$. A discrete-time classic SIRD model is given by
\begin{equation} \label{eq:sird}
\left\{\begin{aligned}
S_{t+1} = &~ 1-I_{t+1}-R_{t+1}-D_{t+1}, \\
I_{t+1} = &~ I_t\left(1+S_tf_\beta(\beta_{t+1})-f_\gamma(\gamma_{t+1})-f_\delta(\delta_{t+1})\right), \\
R_{t+1} = &~ R_t+I_tf_\gamma(\gamma_{t+1}), \\
D_{t+1} = &~ D_t+I_tf_\delta(\delta_{t+1}),
\end{aligned}\right.
\end{equation}
where $f_\beta(\beta_{t+1})$, $f_\gamma(\gamma_{t+1})$, and $f_\delta(\delta_{t+1})$ represent rates of \textit{infection}, \textit{recovery}, and \textit{death} over the period $[t,t+1]$, respectively. Here, we consider all functions $f$'s as sigmoid function\footnote{In general, one can consider inhomogeneous functions $f$'s and modify any of them as $1/(K+e^{-x})$ such that the corresponding rate ranges in $(0,1/K)$ for any constant $K>0$. Although the rates can sometimes exceed 1, especially for the infection rate during the early outbreak, our studies found that using sigmoid function $(K=1)$ yields consistently decent fitting results for US COVID-19 data and thus we will stick with the standard sigmoid function. Note that during the early outbreak of the disease, the data are too less and not reliable.
}, i.e. 
$$
f_\beta(x)=f_\gamma(x)=f_\delta(x)=\sigma(x):=\frac{1}{1+e^{-x}},
$$
such that $\beta_{t+1}$, $\gamma_{t+1}$, and $\delta_{t+1}$ are interpreted as \textit{log-odds} of \textit{infection}, \textit{recovery}, and \textit{death}, respectively. In contrast with most studies that assume constant rates and disregard the range (or positivity) of the rate, this formulation allows us to study the log-odds on the real line and the resulting rates always range in $(0,1)$. It is crucial for later analyses when we introduce mobility controls into the log-odds, since the stochastic modeling (of the log-odds) is now free of constraints, so are the controls. With the data of $\{(I_t,R_t,D_t)\}_{t=0}^T$ and the model \eqref{eq:sird}, we can compute the log-odds as follows: for $t=0,1,\ldots,T-1$,
\begin{equation} \label{eq:rates}
\beta_{t+1}=\hbox{logit}\left(\frac{C_{t+1}-C_t}{I_t(1-C_t)}\right),~ \gamma_{t+1}=\hbox{logit}\left(\frac{R_{t+1}-R_t}{I_t}\right),~ \delta_{t+1}=\hbox{logit}\left(\frac{D_{t+1}-D_t}{I_t}\right),
\end{equation}
where $C_t=I_t+R_t+D_t$ is the total confirmed cases up to time $t$ and $\hbox{logit}(x):=\sigma^{-1}(x)=\ln\left(\frac{x}{1-x}\right)$.


One simple but key observation is that the data we can collect are usually affected by the changing environment. For example, the government and the medical authorities might have made efforts in reducing the infectious cases or deaths by isolation of infected individuals, social distancing, and/or vaccination (if any); see \cite{Cheng2020}. When the vaccination is not available yet, the isolation and social distancing become the most efficient ways to contain the virus situation. Most countries had implemented them when they were aware of seriousness of the disease on their lands; COVID-19 in US is no exception. Hence, when the data are fed into the SIRD model \eqref{eq:sird} or \eqref{eq:rates}, these factors should be taken into account. This paper is focused on fighting COVID-19 by mobility controls (isolation and social distancing) only.
Under the assumption of no vaccination, it is natural to consider that the recovery log-odds $\gamma$ and the death log-odds $\delta$ in \eqref{eq:rates} are specific to the virus and they are not affected by exogenous factors. Throughout this paper, we do not consider the variety of the virus and presume that one country has only one major type of the virus. The infection rate $\beta$, however, is dependent on the mobility of the community.

To join the efforts in fighting COVID-19, Google, a subsidiary of Alphabet Inc., recently published COVID-19 community mobility reports based on data collected from Google products (e.g. Google Maps) to help public health officials to make critical decisions; see \cite{Google}.
The Google mobility indices are also considered as feasible controls as they are directly linked to the government's social distancing policy. One major contribution of this paper is to derive the optimal mobility controls (social distancing). These indices cover mobility changes (compared to the baseline\footnote{The baseline values are computed as the median values of the indices from the \textit{pre-virus} period from Jan 3 to Feb 6, 2020. A zero control vector represents a baseline day. Note that it is defined statistically and there was not be such a day in the past.}) of retail \& recreation (RR), grocery \& pharmacy (GP), parks (PA), transit stations (TS), workplaces (WP), and residential (RE), denoted by a vector $\alpha=(\alpha^{(RR)},\alpha^{(GP)},\alpha^{(PA)},\alpha^{(TS)},\alpha^{(WP)},\alpha^{(RE)})^\top$.

Based on the discussion above, the stochastic modeling of $\beta$ uses mobility indices as independent variables while that of $\gamma$ and $\delta$ does not use any independent variables. To investigate the dynamics of the log-odds, we consider three forms of the dependent variables. Specifically, for $\xi_t=\beta_t$, $\gamma_t$, or $\delta_t$, we consider the dependent variables of the forms: 1) $\xi_t$; 2) $\xi_t-\xi_{t-1}$;  and 3) $\xi_t/\xi_{t-1}-1$. When the residuals are fitted with normality, we are essentially modeling the process as independent normal random variables, the Brownian motion, and geometric Brownian motion for these three forms of dependent variables, respectively. When we attempt to model the relation between the infection rate (log-odds) and the mobility indices, our preliminary studies show that using data of \textit{5-day moving average} for each mobility index produces much better results. It can be understood from an epidemiological perspective that there is an incubation period and the infection is related to the contacts over a certain period. We denote the moving average mobility indices as $\bar{\alpha}=(\bar{\alpha}^{(RR)},\bar{\alpha}^{(GP)},\bar{\alpha}^{(PA)},\bar{\alpha}^{(TS)},\bar{\alpha}^{(WP)},\bar{\alpha}^{(RE)})^\top$. It will be used as the explanatory variables of $\beta$.

\subsection{US COVID-19 data}


Before we present our data analysis, we remark that the COVID-19 data seem to be inaccurate during the early outbreak. For example, in the data collected from Worldometer.info, the total recovery cases would sometimes decrease until Mar 19, 2020. Since $R$ and $D$ are of smaller orders compared to $I$, they have large volatilities and thus when we study the $\{\gamma_t\}$ and $\{\delta_t\}$ processes, we feed only the data starting from Mar 19, 2020. As the controversial data points do not have big impacts on $I$, we still study $\{\beta_t\}$ with the data starting from Mar 1, 2020.

Our first attempt is to fit the three different transformations of $\xi_t=\beta_t$, $\gamma_t$, or $\delta_t$ with normality.
We conduct the Shapiro--Wilk normality tests (see \cite{Shapiro1965}) for the log-odds and the $p$-values are reported in Table \ref{tab:US-SWpval}, where each row represents the testing with the data up to the indicated date. When the $p$-value is large (exceeds a significance level), we cannot reject the normality assumption of the data. We can see that $\{\gamma_t/\gamma_{t-1}-1\}$ can be assumed to be normally distributed, while $\{\delta_t/\delta_{t-1}-1\}$ is comparably normally distributed\footnote{The death rates or cases are likely affected by other factors across time that are not captured with the mobility considered in this paper. However, we still adopt this simple model for simplicity.}. $\{\beta_t-\beta_{t-1}\}$ may also be considered normally distributed marginally.
However,
as we discussed, the infection rate (or log-odds) generally depends on the mobility controls of the country.

\begin{table}[!ht]
	\centering
	\caption{The $p$-values Shapiro--Wilk normality test of the log-odds of the three forms. The largest $p$-value in each cluster of the log-odds is highlighted in bold.}
	\resizebox{\linewidth}{!}{
	\begin{tabular}{c|ccc|ccc|ccc}
		\hline \hline
		Date & $\beta_t$ & $\beta_t-\beta_{t-1}$ & $\beta_t/\beta_{t-1}-1$ & $\gamma_t$ & $\gamma_t-\gamma_{t-1}$ & $\gamma_t/\gamma_{t-1}-1$ & $\delta_t$ & $\delta_t-\delta_{t-1}$ & $\delta_t/\delta_{t-1}-1$ \\
		\hline
		May 31 & $<$1e-04 & \textbf{0.0506} & $<$1e-04 & $<$1e-04 & 0.0009 & \textbf{0.5587} & 0.0012 & 0.0216 & \textbf{0.0355} \\
		Jun 7 & $<$1e-04 & \textbf{0.0275} & $<$1e-04 & $<$1e-04 & 0.0004 & \textbf{0.6304} & 0.0014 & 0.0237 & \textbf{0.0388} \\
		Jun 14 & $<$1e-04 & \textbf{0.0183} & $<$1e-04 & $<$1e-04 & 0.0003 & \textbf{0.6140} & 0.0012 & 0.0116 & \textbf{0.0222} \\
		Jun 21 & $<$1e-04 & \textbf{0.0202} & $<$1e-04 & $<$1e-04 & 0.0001 & \textbf{0.6895} & 0.0012 & 0.0047 & \textbf{0.0102} \\
		\hline \hline
	\end{tabular}
	}
	\label{tab:US-SWpval}
\end{table}


Next, we regress each transformation of the infection log-odds on the 5-day moving average mobility indices $\bar{\alpha}=(\bar{\alpha}^{(RR)},\bar{\alpha}^{(GP)},\bar{\alpha}^{(PA)},\bar{\alpha}^{(TS)},\bar{\alpha}^{(WP)},\bar{\alpha}^{(RE)})^\top$, i.e. we consider the regression model $\hbox{LHS}_{t+1}=c_0+c^\top \bar{\alpha}_t+\epsilon_{t+1}$ for $t=1,\ldots,T-1$, where $\hbox{LHS}_{t+1}=\beta_t$, $\beta_t-\beta_{t-1}$ or $\beta_t/\beta_{t-1}-1$, $c_0$ and $c$ are regression coefficients, and $\epsilon_{t+1}$ is the error term at time $t+1$. The regression results for each period are presented in Table~\ref{tab:US-fit}\footnote{For the fittings of using $\beta_t-\beta_{t-1}$ or $\beta_t/\beta_{t-1}-1$ as dependent variable, we found that truncation of the early problematic data points (first 20 data) improve the model fit. However, the truncation does not improve the fitting of using $\beta_t$ as dependent variable. Table~\ref{tab:US-fit} presented the best results for each of them and using $\beta_t$ is still the winner.}.
With the mobility indices as independent variables, the fittings are greatly improved and promising. The best models consistently use the $\beta_t$ as dependent variable and in this case, all the coefficients of determination ($R^2$) are larger than 0.84 while the residuals pass the Shapiro--Wilk normality test with any significance level larger than 9.8\%.
\begin{table}[!ht]
	\caption{Regression table of fitting the infection log-odds with moving average mobility indices. In parentheses under the coefficient estimates are their standard errors. The last column reports the $p$-value of the Shapiro--Wilk normality test of the residuals. The codes of significance are [0 `***' 0.001 `**' 0.01 `*' 0.05 `\#' 0.1 ` ' 1]. The date in header represents the fitting date.
	 The best fitting in each period is highlighted in bold.}
	\centering
	\resizebox{\linewidth}{!}{
	\begin{tabular}{c|ccccccc|c|c}
		\hline \hline
		LHS & (Intercept) & $\bar{\alpha}^{(RR)}$ & $\bar{\alpha}^{(GP)}$ & $\bar{\alpha}^{(PA)}$ & $\bar{\alpha}^{(TS)}$ & $\bar{\alpha}^{(WP)}$ & $\bar{\alpha}^{(RE)}$ & $R^2$ & $p$-value \\
		\hline
		\multicolumn{10}{c}{May 31} \\
		\hline
		\multirow{2}{*}{\bm{$\beta_t$}} & \textbf{0.5824*}  & \textbf{-20.6616***} & \textbf{5.2851***} & \textbf{-2.0522**} & \textbf{32.4049***} & \textbf{3.4093}   & \textbf{34.9708***} & \multirow{2}{*}{\textbf{0.8506}} & \multirow{2}{*}{\textbf{0.1706}} \\
		& (0.2527) & (3.4974) & (1.2411)  & (0.7075)  & (5.0816)   & (4.3353) & (7.9842) &  &  \\
		\hline
		\multirow{2}{*}{$\beta_t-\beta_{t-1}$} & 0.1378   & 3.5887*** & -1.5294\# & -0.5604\# & -2.5342    & -1.9667  & -6.3625** & \multirow{2}{*}{0.3470} & \multirow{2}{*}{0.0413} \\
		 & (0.5303) & (1.0015)    & (0.8152)  & (0.3157)  & (2.3434)   & (1.2252) & (1.8915) &  &  \\
		\hline
		\multirow{2}{*}{$\beta_t/\beta_{t-1}-1$} & 0.5811   & -3.7691***  & 1.5588*   & -0.0191   & 4.9242**   & 0.3565   & 3.9427** & \multirow{2}{*}{0.5170} & \multirow{2}{*}{$<$1e-04} \\
		& (0.3865) & (0.7298)    & (0.5940)  & (0.2301)  & (1.7077) & (0.8928) & (1.3784) &  &  \\
		\hline \hline
		\multicolumn{10}{c}{Jun 7} \\
		\hline
		\multirow{2}{*}{\bm{$\beta_t$}} & \textbf{0.4205\#} & \textbf{-18.0781***} & \textbf{4.9546***} & \textbf{-1.9228**} & \textbf{28.8964***} & \textbf{5.2908} & \textbf{36.7253***} & \multirow{2}{*}{\textbf{0.8549}} & \multirow{2}{*}{\textbf{0.2183}} \\
		& (0.2288) & (3.2160) & (1.1862) & (0.6214) & (4.4397) & (3.6844) & (7.6324) &  &  \\
		\hline
		\multirow{2}{*}{$\beta_t-\beta_{t-1}$} & 0.1514 & 3.5472*** & -1.5934* & -0.5584\# & -2.4748 & -1.9576\# & -6.3926*** & \multirow{2}{*}{0.3447} & \multirow{2}{*}{0.0533} \\
		& (0.5216) & (0.8950) & (0.7875) & (0.3061) & (2.2280) & (1.0967) & (1.8285) &  &  \\
		\hline
		\multirow{2}{*}{$\beta_t/\beta_{t-1}-1$} & 0.5617 & -3.425*** & 1.5624** & -0.0369 & 4.487** & 0.4889 & 3.9627** & \multirow{2}{*}{0.5065} & \multirow{2}{*}{$<$1e-04} \\
		 & (0.3713) & (0.6371) & (0.5606) & (0.2179) & (1.5859) & (0.7806) & (1.3016) &  &  \\
		\hline \hline
		\multicolumn{10}{c}{Jun 14} \\
		\hline
		\multirow{2}{*}{\bm{$\beta_t$}} & \textbf{0.3180} & \textbf{-16.1974***} & \textbf{4.5635***} & \textbf{-1.7821**} & \textbf{26.6568***} & \textbf{5.8699} & \textbf{36.6619***} & \multirow{2}{*}{\textbf{0.8526}} & \multirow{2}{*}{\textbf{0.1589}} \\
		& (0.2268) & (3.1344) & (1.1792) & (0.5747) & (4.3151) & (3.6084) & (7.6185) &  &  \\
		\hline
		\multirow{2}{*}{$\beta_t-\beta_{t-1}$} & 0.0273 & 3.7934*** & -1.3501** & -0.5879 & -3.2047** & -1.8478 & -6.6743** & \multirow{2}{*}{0.3768} & \multirow{2}{*}{0.022} \\
		 & (0.4811) & (0.8261) & (0.6988) & (0.2917) & (2.0765) & (1.0457) & (1.7541) &  &  \\
		\hline
		\multirow{2}{*}{$\beta_t/\beta_{t-1}-1$} & 0.6694* & -3.2298*** & 1.3405** & -0.0853 & 4.6834** & 0.3634 & 3.8151** & \multirow{2}{*}{0.4992} & \multirow{2}{*}{$<$1e-04} \\
		 & (0.3417) & (0.5869) & (0.4964) & (0.2072) & (1.4751) & (0.7428) & (1.2460) &  &  \\
		\hline \hline
		\multicolumn{10}{c}{Jun 21} \\
		\hline
		\multirow{2}{*}{\bm{$\beta_t$}} & \textbf{0.3282} & \textbf{-16.9957***} & \textbf{4.4961***} & \textbf{-1.4419**} & \textbf{28.5379***} & \textbf{3.7798} & \textbf{34.6980***} & \multirow{2}{*}{\textbf{0.8413}} & \multirow{2}{*}{\textbf{0.0988}} \\
		& (0.2313) & (3.0850) & (1.2010) & (0.5302) & (4.3214) & (3.5796) & (7.7061) &  &  \\
		\hline
		\multirow{2}{*}{$\beta_t-\beta_{t-1}$} & 0.0965 & 3.6512*** & -1.6301* & -0.5173\# & -2.7958 & -1.6731 & -6.0067** & \multirow{2}{*}{0.3534} & \multirow{2}{*}{0.0388} \\
		 & (0.4690) & (0.8019) & (0.6658) & (0.2959) & (1.9596) & (1.0068) & (1.7752) &  &  \\
		\hline
		\multirow{2}{*}{$\beta_t/\beta_{t-1}-1$} & 0.6708* & -3.0629*** & 1.357** & -0.1323 & 4.5882*** & 0.2053 & 3.5132** & \multirow{2}{*}{0.4871} & \multirow{2}{*}{$<$1e-04} \\
		 & (0.3170) & (0.5420) & (0.4500) & (0.2000) & (1.3245) & (0.6805) & (1.1998) &  &  \\
		\hline \hline
	\end{tabular}
	\label{tab:US-fit}
}
\end{table}

\begin{table}[!ht]
	\centering
	\caption{The parameter estimates of the model in \eqref{eq:ratesdynamics} for different periods using historical data.}
	\begin{tabular}{c|cc|cc|c}
		\hline \hline
		Date & $\mu_\gamma$ & $\sigma_\gamma$ & $\mu_\delta$ & $\sigma_\delta$ & $\sigma_\beta$ \\
		\hline
		May 31 & 0.0107	& 0.1987 & 0.0058 & 0.0432 & 0.5036 \\
		Jun 7 & 0.0178	& 0.1996 & 0.0062 & 0.0433 & 0.4988 \\
		Jun 14 & 0.0173 & 0.1963 & 0.0061 & 0.0444 & 0.5032 \\
		Jun 21 & 0.0176 & 0.1919 & 0.0061 & 0.0451 & 0.5166 \\
		\hline \hline
	\end{tabular}
	\label{tab:US-par}
\end{table}

Therefore, for the COVID-19 process in US, a suitable stochastic model for the log-odds is proposed as follows: for $t=1,\ldots,T-1$,
\begin{equation} \label{eq:ratesdynamics}
\beta_{t+1}=c_0+c^\top \bar{\alpha}_t+\sigma_\beta Z_{t+1}^{\beta}, \quad \gamma_{t+1}=\gamma_t(1+\mu_\gamma+\sigma_\gamma Z_{t+1}^{\gamma}), \quad
\delta_{t+1}=\delta_t(1+\mu_\delta+\sigma_\delta Z_{t+1}^{\delta}),
\end{equation}
where $\bar{\alpha}_t=\frac{1}{5}\sum_{t-4}^t\alpha_t$ and $Z_{t+1}^{\beta},Z_{t+1}^{\gamma},Z_{t+1}^{\delta}$ are independent and identically distributed (iid) $N(0,1)$ random variables for $t=1,\ldots,T-1$. The intercept $c_0$ and the coefficients $c\in \mathbb{R}^6$ can be found in Table~\ref{tab:US-fit} for the fittings on different dates and other parameters (mean $\mu$ and standard deviation $\sigma$) can be easily estimated with historical data as given in Table~\ref{tab:US-par}. From Table~\ref{tab:US-fit} and Table~\ref{tab:US-par}, we can see that the parameter estimates are quite stable for the whole month of June. The intercept $c_0$ and the coefficient corresponding to $\bar{\alpha}^{(WP)}$ are sometimes statistically insignificant but we still keep them for the consistency of the model and the purpose of capturing all mobility effects on economics and controls. The interpretation of the regression coefficients in Table~\ref{tab:US-fit} requires extra cautions as the mobility indices are highly correlated with each other, a phenomenon known as multicollinearity. However, as we just use the regression model for predicting the drift of $\beta_{t+1}$ in \eqref{eq:ratesdynamics}, the multicollinearity does not bother the prediction.

With the fitted model \eqref{eq:ratesdynamics}, we can conduct simulations to evaluate the effectiveness of the government's responses to COVID-19. Fixing the future mobility indices as the median ones for each period corresponding to a social distancing policy, we find that only school closure in US will still lead to massive infectious cases and deaths, while the COVID-19 can be contained by the end of 2020 with school and workplace closures. The results are presented in the Appendix~\ref{appendix:1}.

\section{Learning COVID-19 driven economic value from financial data} \label{sec:econ}

Although restricting social mobility has positive effect in containing the disease, there are opportunity costs from reducing the mobility. At least, nobody is happy with the lockdown of the economy. The aggregated economic cost should reflect the risk from economic downturn, the spread of the disease, and the social dissatisfaction. Therefore, the optimal mobility control strikes the best balance between consumption and health of the population. A similar considerations in \cite{Hall2020} investigates the trade-off between the consumption and COVID-19 deaths.  
The novelty of our approach is to learn the aggregated economic cost as a functional of SIRD and mobility indices through the financial market data.

Since the outbreak of COVID-19, the US government has made different responses to the coronavirus situation over different periods; see \cite{Cheng2020,OWID}. In supplementary appendix, we focus on some past representative mobility indices and conduct simulation studies to investigate their implications.
 However, the government is not limited by the past responses and can launch some new or combined social distancing rules to control the community mobility effectively. To find the optimal mobility control policy, we need to first identify the trade-off among the COVID-19 statistics, the mobility controls, and the target of our interest. Some studies concerning about the healthcare burden set the cost functional as the healthcare costs. A more recent study in \cite{Benzell2020} quantifies the lockdown decision as a trade-off between infection risks and benefits, by using multiple indicators coming from both economic and consumers' contributions. We share the view that during a pandemic crisis like COVID-19, the financial market is dominated by the news about the COVID-19 as it affects many businesses. Especially when the country is locked down, the economics and most businesses in the country will be lashed. It characterizes a trade-off between the mobility controls and the economics of the country.

In this subsection, we explore the relationships among the COVID-19 statistics, the mobility controls, and the market index. The market index, chosen as S\&P 500 index for the US in this paper, is used to reflect the economics of the US. Once the relationships are identified, we postulate that the economic health is reflected through the stability of the market during the COVID-19 crisis period prior to any economic stimulus. Efficient mobility control policy attempts to achieve public health goals (such as target infection rate less than 1) with minimum cost on economic health. To conceptualize the efficiency, we search for the optimal mobility control policy set that minimizes the acquired cost functional incorporating both public and economic health.


\subsection{S\&P 500 index price data}

The S\&P 500 index prices are downloaded from Yahoo!Finance. We conduct an ex-post regression of the daily S\&P 500 index closing prices ($\hbox{SPX}_t$) on the COVID-19 statistics and mobility indices. Specifically, we postulate that
$$
\hbox{SPX}_t=\kappa_0+\kappa^\top \alpha_t+\kappa_I I_t+\kappa_R R_t+\kappa_D D_t+\epsilon^{SPX}_t \label{SPX}
$$
for $t=1,\ldots,T$, where $\kappa_0$, $\kappa\in\mathbb{R}^6$, $\kappa_I$, $\kappa_R$, and $\kappa_D$ are regression coefficients and $\epsilon^{SPX}_t$ is the error term at time $t$. Note that in the independent variables, we use $\alpha$ instead of the moving average one $\bar{\alpha}$ because the market's reaction to the community mobility is immediate. The regression results are reported in Table~\ref{tab:SPXreg}. Each fitting is based on the daily data (working days) from Mar 9 to the fitting date indicated in the table.

\begin{table}[!ht]

\caption{Regression table of fitting S\&P 500 index prices with COVID-19 statistics and mobility indices in US. In parentheses under the coefficient estimates are their standard errors. The R-squared and the $p$-value of the Shapiro--Wilk normality test of the residuals are reported next to the fitting dates in the headers . The codes of significance are [0 `***' 0.001 `**' 0.01 `*' 0.05 `\#' 0.1 ` ' 1].}

\centering

\resizebox{\linewidth}{!}{

\begin{tabular}{cccccccccc}

\hline \hline

(Intercept) & $\alpha^{(RR)}$ & $\alpha^{(GP)}$ & $\alpha^{(PA)}$ & $\alpha^{(TS)}$ & $\alpha^{(WP)}$ & $\alpha^{(RE)}$ & $I$ & $R$ & $D$ \\

\hline

\multicolumn{10}{c}{\textbf{May 31} ($R^2=0.8903$, $p$-value$=0.6291$)} \\

\hline

2919.7*** & 2.5 & -832.7*** & -131.1 & 1847.4 & -64.0 & 830.4 & 5.958e05*** & 3.943e05* & -6.493e06** \\

(96.1) & (722.6) & (235.9) & (197.6) & (1049.7) & (824.2) & (918.0) & (1.374e05) & (1.485e05) & (1.972e06) \\

\hline \hline

\multicolumn{10}{c}{\textbf{Jun 7} ($R^2=0.9075$, $p$-value$=0.3995$)} \\

\hline

2949.2*** & -71.5 & -811.1*** & -233.6 & 2202.0* & -288.6 & 790.6 & 6.508e05*** & 5.578e05*** & -7.668e06*** \\

(89.1) & (667.8) & (230.6) & (182.0) & (978.5) & (793.7) & (894.4) & (1.304e05) & (1.128e05) & (1.815e06) \\

\hline \hline

\multicolumn{10}{c}{\textbf{Jun 14} ($R^2=0.9039$, $p$-value$=0.3899$)} \\

\hline

2917.0*** & 601.5 & -943.5*** & -314.3 & 1656.3 & -380.7 & 855.7 & 4.925e05*** & 4.008e05*** & -5.361e06** \\

(92.9) & (655.3) & (243.2) & (190.9) & (1011.5) & (841.8) & (945.1) & (1.276e05) & (9.899e04) & (1.741e06) \\

\hline \hline

\multicolumn{10}{c}{\textbf{Jun 21} ($R^2=0.8979$, $p$-value$=0.4141$)} \\

\hline

2875.8*** & 1108.1\# & -1100.5*** & -301.0 & 1001.4 & -179.1 & 1195.2 & 2.772e05* & 1.957e05* & -2.217e06 \\

(91.4) & (623.6) & (246.3) & (193.9) & (982.4) & (862.5) & (963.2) & (1.111e05) & (7.793e04) & (1.470e06) \\

\hline \hline

\end{tabular}

}

	\label{tab:SPXreg}
\end{table}
From Table~\ref{tab:SPXreg}, we can see that the fittings are all good as all $R^2$ are higher than 0.89 and the residuals pass the Shapiro--Wilk normality test with any significance level larger than 38\%.
It shows that during the COVID-19 crisis, the COVID-19 statistics and the mobility indices have high explanatory power for the S\&P 500 index price. Even some independent variables are insignificant, we choose to keep all the independent variables because we look for a high explanatory power and in our preliminary studies on the control problem below, eliminating some mobility controls will lead to the extreme values of the eliminated mobility controls in optimality. Again, we need to be cautious in interpreting the coefficients as there is also multicollinearity. However, it does not affect our use of the regression model as the proxy for the COVID-19 driven economic values learnt from the S\&P 500 index price.

\section{Efficient social distancing}\label{Section4}

When we aim to reduce an integrated risk of public health and economic health through social distancing, it can be conceptually written as
\begin{equation}
\min_{\alpha} ~\{(\hbox{public health risk}) + \lambda (\hbox{economic health risk})\}, \label{concept}
\end{equation}
where $\alpha$ is a mobility control and $\lambda \ge 0$ reflects the economic risk aversion. When $\lambda =0$, the policymaker only cares about the public health even at an unlimited economic cost. In other words, a policymaker concerning the economic health should take a positive $\lambda$. However, an extremely large $\lambda$ means almost ignoring the public health issue. 
For any fixed $\lambda \ge 0$, if the conceptual problem \eqref{concept} can be solved, then the optimal mobility control $\alpha^*(\lambda)$ is called an efficient social distancing policy (ESDP). We define the efficient social distancing frontier (ESDF) as the set of all ESDPs: ESDF = $\{ \alpha^*(\lambda): \lambda \ge 0\}$. 

The conceptual framework \eqref{concept} and the ESDF constitute an epidemic-economic analogue to the mean-variance portfolio theory by \cite{Markowitz1952}, a Nobel memorial prize work. The key difference is that portfolio theory focuses on mean and variance of a single random output (the investor's wealth) whereas the efficient social distancing problem \eqref{concept} has two different random outputs associated with the epidemic and economic issues, respectively.

An obvious COVID-19 public health target is to contain infection by social distancing. In a statistical perspective, we concern about the probability of increasing in the infection rate $\sigma(\beta_t)$ with a social distancing policy against the probability of decreasing. The ratio between the two probabilities is essentially the odds of infection. In Section \ref{sec:stomodel}, we have spent a lot of effort to statistically learn how the log-odds of infection $\beta_t$ depends on the stochastic epidemic model with mobility controls using an empirical dataset. We, therefore, set the public health target as minimizing the expected log-odds of infection: $\min E[\beta_t],$ at a future time $t$. In other words, the public health risk is set as the expected future log-odds of inflection. In fact, the log-odds of infection is an increasing function of the infection rate but it is more sensitive to the change in social distancing policy, facilitating the numerical optimization efficiency.

We model the economic health as the stability of the economic value associated with the epidemic model and social distancing policy. With the predicted S\&P 500 index price over the COVID-19 crisis period, we statistically learn the economic value in SPX as
\begin{equation} \label{eq:SPX}
\widehat{\hbox{SPX}}_t=\hat{\kappa}_0+\hat{\kappa}^\top \alpha_t+\hat{\kappa}_{I}I_t+\hat{\kappa}_{R}R_t+\hat{\kappa}_{D}D_t,
\end{equation}
where $\hat{\kappa}_0$, $\hat{\kappa}$, $\hat{\kappa}_{I}$, $\hat{\kappa}_{R}$, and $\hat{\kappa}_{D}$ appear in Table~\ref{tab:SPXreg}. The economic health risk is the tracking error between the COVID-19 driven economic growth and a reasonable growth. Mathematically, the tracking error reads
$$\left(\frac{\widehat{\hbox{SPX}}_t - \widehat{\hbox{SPX}}_0}{\widehat{\hbox{SPX}}_0} - r_t\right)^2 \propto \left(\widehat{\hbox{SPX}}_t-\hbox{Target}_t\right)^2,$$
where $r_t$ is the target growth rate and Target$_t = \widehat{\hbox{SPX}}_0(1 + r_t)$ is the target economic value. A reasonable growth rate $r_t$ could be the risk-free interest rate, the inflation rate, or GDP growth rate etc. However, when $t$ is small for the case of near future, all of the target rates are close to zero. In fact, minimizing the expected tracking error with $r_t= 0$ aims to reducing the risk of significant economic downturn without emphasising on the economic growth.

Therefore, our conceptual framework of integrating public health and economic health in \eqref{concept} is modelled through the (running) cost functional:
\begin{equation} \label{eq:cost}
c(t,X_t, \lambda;\alpha_t)=\beta_t + \lambda (\widehat{\hbox{SPX}}_t-\hbox{Target}_t)^2,
\end{equation}
where $X_t:=(I_t,R_t,D_t,\beta_t,\gamma_t,\delta_t)^\top$. In reality, a policymaker seldom considers the effect on a single future day but rather a period of future time $h$.  For instance, if the policymaker considers the projection for 5 working days in a week, then $h=5$, which is also the choice of our numerical study. Therefore, we formulate and quantify the efficient social distancing problem \eqref{concept} into a stochastic control problem:
\begin{equation} \label{eq:prob}
\begin{aligned}
\min_{\alpha_s \in \mathcal{A}_s,~s=t,\ldots,t+h-1} & \qquad J(t,X_t;\{\alpha_s \}_{s=t}^{t+h-1}):=\mathbb{E}\left[\sum_{s=t+1}^{t+h}c(s,X_{s}, \lambda;\alpha_{s-1})\right] \\
\hbox{s.t.} & \qquad \{X_s\}_{s=t+1}^{t+h} \hbox{ satisfies \eqref{eq:sird} and \eqref{eq:ratesdynamics}}, ~\widehat{\hbox{SPX}}_t \hbox{ satisfies }\eqref{eq:SPX},
\end{aligned}
\end{equation}
where the cost functional $c$ is given in \eqref{eq:cost}, and $\mathcal{A}_s$ is the feasible set of mobility controls determined at time $s$.  As discussed before, the mobility indices are highly correlated with each other. We characterize $\mathcal{A}_s$ using principal component analysis (PCA). This enhances the searching efficiency of the computation without sacrificing reasonable feasible controls.  The idea is to use the historical data of the mobility indices to determine the transformation from indices $\alpha$ to their PCs:  $P=\hat{A}\alpha$, where $\hat{A}\in \mathbb{R}^{6\times6}$ is a matrix to be estimated. We then compute the historical PCs, $P\in \mathbb{R}^6$, and determine the lower and upper bounds of each entry of the historical PCs, namely $\hat{L}\in \mathbb{R}^6$ and $\hat{U}\in \mathbb{R}^6$, respectively. The definition of $\alpha$ requires that each entry of the $\alpha$ falls in the range of $[-1,1]$. Let $\hat{A}_t$, $\hat{L}_t$, and $\hat{U}_t$ be the estimates using the data up to time $t$. Then the feasible set $\mathcal{A}_s$ at time $s \in [t,t+h-1]$ is given by
\begin{equation} \label{eq:fea}
\mathcal{A}_s:=\{ \alpha_s(w_s) : \mathbb{R}^{30} \rightarrow \mathbb{R}^6 ~|~\hat{L}_t\le_e\hat{A}_t\alpha_s\le_e\hat{U}_t \hbox{ and } -\mathbf{1}\le_e \alpha_s \le_e \mathbf{1}\},
\end{equation}
where $\le_e$ is the entry-wise less-than-or-equal-to operator and  $\mathbf{1}=(1,\ldots,1)^\top \in\mathbb{R}^6$. Note that the feasible set \eqref{eq:fea} is an approximation of the full feasible set. However, our numerical studies show that it works well in practice and the ESDPs found are close to some realized mobility indices.

To incorporate the PC constraints on $\alpha_s$ into the objective function, we use a Lagrangian approach such that the revised cost is
\begin{equation}
\overline{c}^\lambda_{s}  = c_{s} + \sum_{i=1}^{6}\overline{\lambda}_i \left[(\hat{A}_t\alpha_s -\hat{L}_t)_i^+  -(\hat{A}_t\alpha_s-\hat{U}_t)_i^-\right]^2, \quad t \le s \le t+h-1,\label{revise}
\end{equation}
where $(a)_i$ denotes the $i^{\text{th}}$ entry of the vector $a$, and $\overline{\lambda}_i$ is a penalty coefficient for $i = 1, 2, \ldots, 6$. Using the available dataset, cross-validation is performed to select $\overline{\lambda}_i$ to ensure that the optimal controls we learned strictly satisfy all of the constraints. Cross-validation is a classic and well-received machine learning and statistical learning technique to identify the most useful opportunity set. 

We solve \eqref{eq:prob} numerically by the deep learning framework in \cite{Tsang2020} . The implementation procedure is detailed in Appendix~\ref{deep}.

\section{Results and discussion}\label{sec:result}

\begin{figure}[!ht]
	\centering
	\begin{minipage}{0.45\textwidth}
		\caption{Efficient social distancing frontier}\label{fig3}
		\includegraphics[width=1\textwidth]{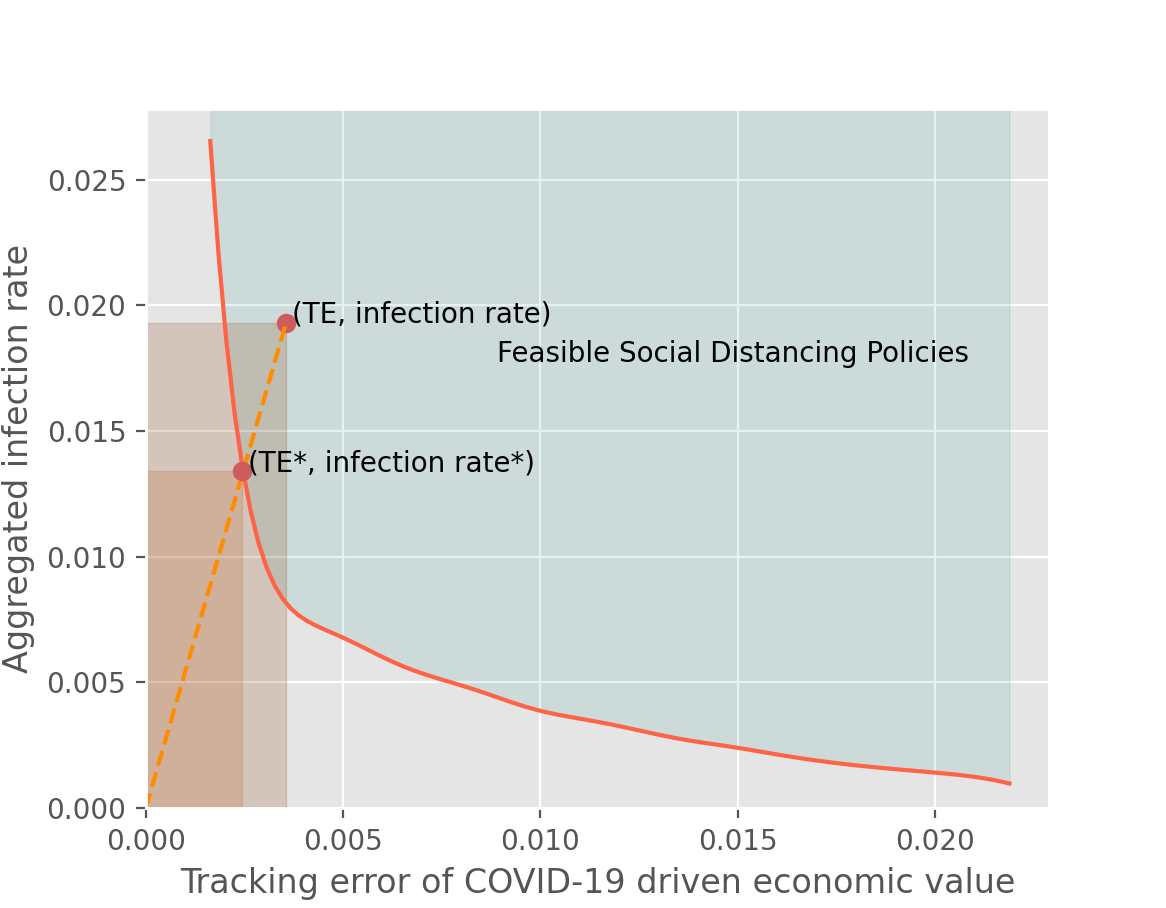} \nonumber\\ 
		\centering
	\end{minipage}
	\begin{minipage}{0.45\textwidth}
		\caption{ESDF with different increment rates}\label{fig4}
		\includegraphics[width=1\textwidth]{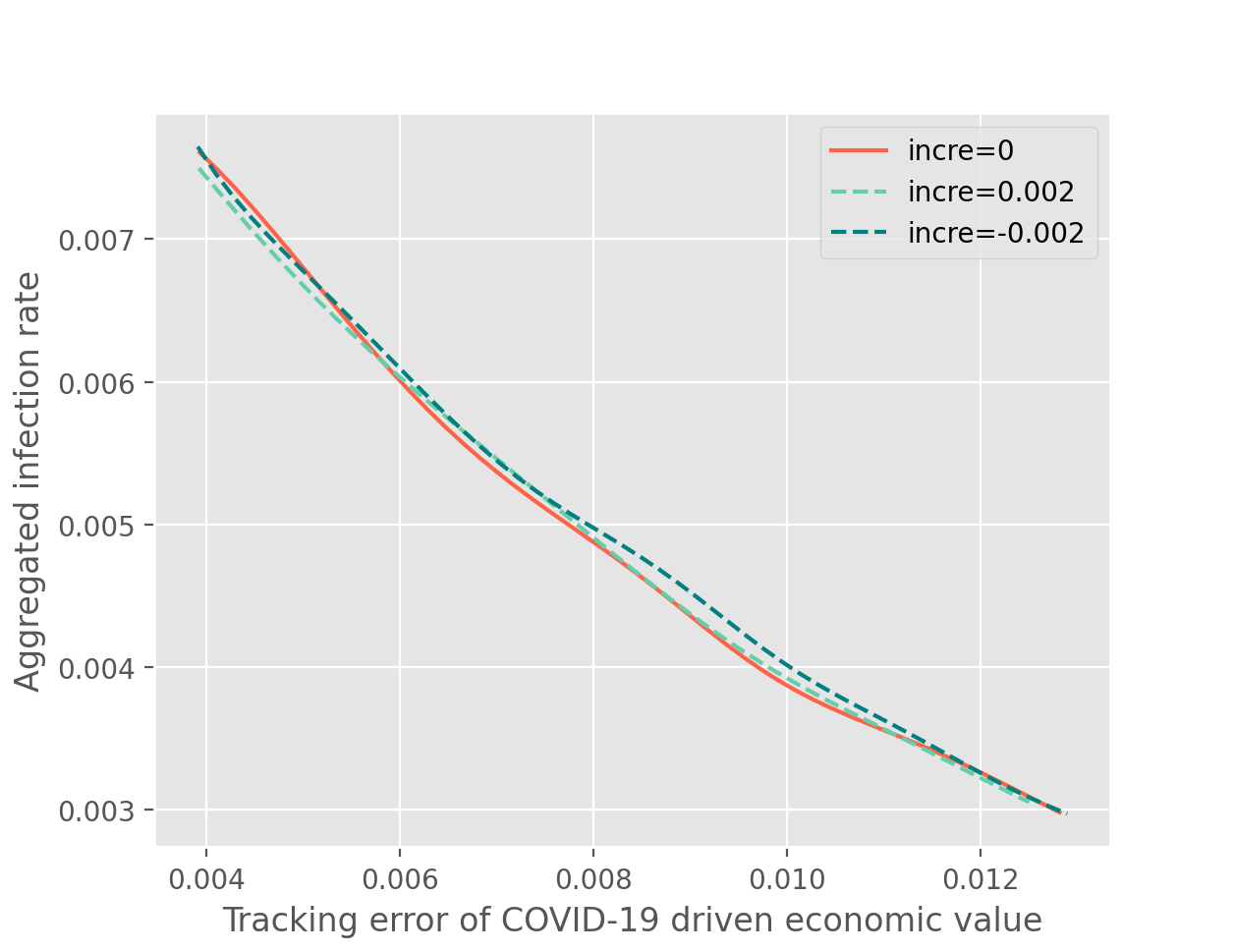} \nonumber\\ 
		\centering
	\end{minipage}
\end{figure}

Figure~\ref{fig3} presents the ESDF for the week containing Jun 29, 2020 by setting the target daily increment rate as 0. In other words, we want the COVID-19 driven economic value found in the S\&P 500 index to stay near the initial economic level calibrated on Jun 26 2020, in the upcoming week. To draw a better interpretation, the $y-$axis of Figure~\ref{fig3} is the aggregated infection rate which is transformed from the log-odds of infection as $\sigma(\sum_{h=1}^5\beta_{t+h}/5)$ whereas the $x-$axis is the tracking error of the COVID-19 driven economic value. To match the order of magnitude between the infection rate and the tracking error, we report the tracking error (TE) in terms of root mean square error (RMSE) in returns. Specifically, 
$$\hbox{TE} = \sqrt{\frac{1}{n}\sum_{t=1}^n \left(\frac{\widehat{\hbox{SPX}}_t-\widehat{\hbox{SPX}}_0}{\widehat{\hbox{SPX}_0}} - r_t\right)^2}.$$
In Figure~\ref{fig3}, the shaded area above the ESDF is the region of epidemic-economic outcomes from the set of feasible social distancing strategies. The lower bullet shape ESDF curve is drawn by varying the $\lambda$. A sensitivity analysis of the ESDF against the target rate $r_t$ is shown in Figure~\ref{fig4}. It is clear that a reasonable choice of $r_t$ does not materially affect the ESDF, which consistently maintains a lower bullet-shape curve. 

To better describe our empirical results, we introduce an empirical quantity for measuring the efficiency of a social distancing policy. For any feasible social distancing policy illustrated as a point inside the feasible set in  Figure~\ref{fig3}, we can search for the intersection point $(\hbox{TE}^*, \hbox{infection rate}^*)$ between the ESDF and the straight line connecting the feasible point and the origin. The intersection point is referred to as the benchmarking ESDP point. For any feasible social distancing policy, we define the public-economic risk area (PERA) as the product between the infection rate and the TE, illustrated as the large rectangle in Figure~\ref{fig3}. Similarly, we measure the benchmarking ESDP's PERA (BEPERA) in the same manner, illustrated as the small rectangle. Hence, the efficiency ratio (ER) is defined as 
$$\hbox{ER} = \frac{\hbox{BEPERA}}{\hbox{PERA}}= \frac{\hbox{TE}^*\times \hbox{infection rate}^*}{\hbox{TE}\times \hbox{infection rate}}\in [0,1].$$
The greater the ER the higher the efficiency of a feasible social distancing policy. Clearly, ESDP has ER equal to 100\%.
\begin{figure}[!ht]
	\centering
	\caption{US mobility index controls. The shaded area depicts the range of average efficient mobility controls over the validation data set. The solid line represent the historical mobility.}
	\includegraphics[width=\linewidth]{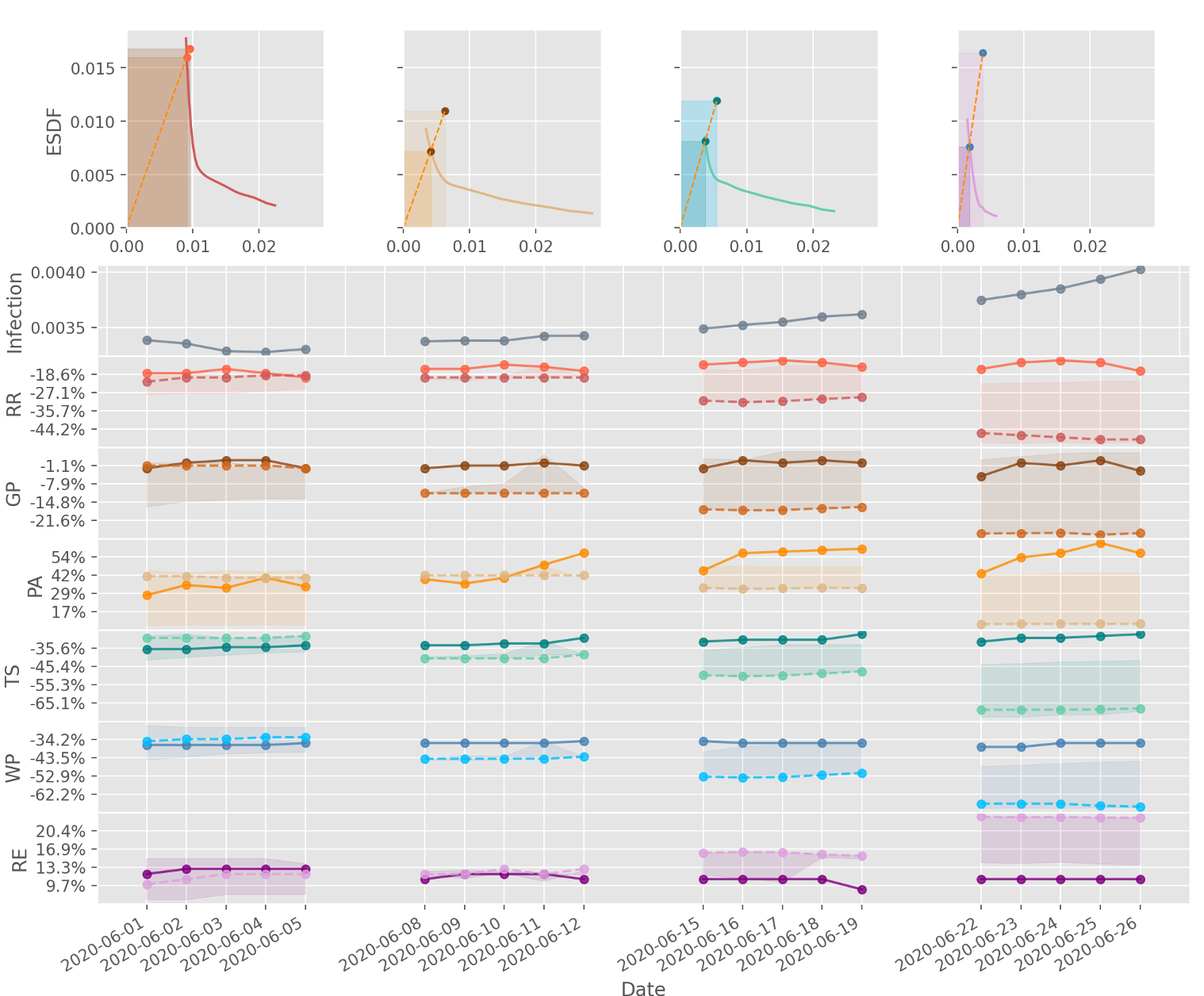}
	\label{fig2}
\end{figure}
 Figure~\ref{fig2} displays the ESDPs for the validation dataset from June 8, 2020 to June 26, 2020. We set $r_t= 0$ and compute the ESDP for $\lambda \in (0.001, 0.1)$. We highlight that the computed ESDPs all satisfy the PC constraints on the control. We aim to visualize the efficiency of historical social distancing by benchmarking against our ESDPs in Figure~\ref{fig2}. The top panel of the graph shows the weekly ESDFs (the curves), historical epidemic-economic points (corresponding to historical social distancing policies) and the benchmarking ESDPs, over four weeks in June, 2020. In this way, we can compute the ERs for these four weeks. We recognize by inspection and, in fact, the ER number also shows that the first week's policy is the most efficient one among the four. In fact, the ER
numbers for the four weeks are 91.13\%, 43.77\%, 46.99\% and 21.92\%, respectively. The ER trend
implies that the historical social distancing is losing its efficiency in this particular month.

The second panel of  Figure~\ref{fig2} shows the trend of historical active infections per population over the four weeks in June, 2020.  We recognize a mild decreasing trend in the first week. However, the increasing trend begins in the second week and the number of active infected cases keeps increasing in the rest of the month. This seems related to the efficiency of the social distancing scheme.

The third panel of  Figure~\ref{fig2} reports the historical mobility indices, namely RR, GP, PA, TS, WP and RE, and the ESDPs. Solid lines are the realized historical mobility indices in the dataset. The historical indices are typically more volatile over the weekends than weekdays. Therefore, we only report the results for weekdays (working days). Although all of the historical mobility indices are seemingly stable over the month, we recognize slight increases in RR and TS, and a clear rising trend in PA in the last three weeks when compared to the situation in the first week, indicating certain level of relaxation in mobility. The shaded area in each mobility index represents the range of ESDPs for $\lambda\in (0.001, 0.1)$. The dash lines are those from the recommended ESDP, which is obtained as follows. We use the mobility indices from the last Friday and compute the (TE, infection rate) of the current week from the training sample. Then, we find the $\lambda$ by which the ESDP offers the same TE. Such an ESDP is our recommended ESDP. The intent of doing so aims to maintain the same level of economic risk while pulling down the infection rate to the efficient level. It can be seen from Figure~\ref{fig2} that the first week's policy is very close to the recommended ESDP. In the following weeks, the recommended ESDP suggests to reduce mobilities in all areas except for RE. It seems that the data-driven ESDP computed from our algorithm provides a reliable reference for social distancing policymakers.

\section{Conclusion}\label{sec:con}

This paper leverages the community mobility reported by
Google to model the infection process of COVID-19 and the economy of a country. Using the US COVID-19 data, we found that the fittings of a stochastic SIRD model are significantly improved with mobility indices and during the pandemic crisis, the explanatory power of the mobility indices on the market index (S\&P 500 index) is high. Our stochastic modeling of US COVID-19 allows us to simulate the possible scenarios given certain mobility in the future and thus we can verify the effectiveness of a social distancing policy. Moving forward, this paper also provides efficient social distancing policies by solving a dynamic stochastic control problem, where we initiate an objective integrating economic health and public health. Such a realistic problem is posed with analytical challenges while we manage to solve it with a deep-learning approach. The results are informative and the policymakers can refer to the model's suggestion to design the social distancing policy.



\begin{appendix}
\section{The responses to COVID-19 and their implications in simulation studies}\label{appendix:1}

The SIRD model \eqref{eq:sird} with stochastic processes of \eqref{eq:ratesdynamics} can be used to describe the evolution of the infectious disease in the country. In this subsection, we conduct simulation studies to analyze some scenarios of COVID-19 in US.

Note that the daily mobility indices of a country are random and different indices are highly correlated with each other. It is very difficult to fix the community mobility $\alpha$ as an arbitrary vector. In this subsection, we investigate the real historical data of the mobility and investigate how certain real mobility affect the evolution of the COVID-19. We are focused on some representative mobility in the past. For different measures of social distancing and restriction of public gatherings by the government, they will be reflected on the community mobility. Since the outbreak of COVID-19, the US government has made different responses to the coronavirus situation over different periods; see \cite{Cheng2020,OWID}. Based on the government's responses, we consider four periods and we take the median mobility indices as the representative community mobility in each period; see Table~\ref{tab:medmob}.

\begin{table}[!ht]
	\caption{The median mobility indices over different periods with different major measures by the government.}
	\resizebox{\linewidth}{!}{
	\begin{tabular}{r|c|l}
		\hline \hline
		Period & Government's responses & Median mobility indices \\
		\hline
		Jan 3 -- Feb 6 & Baseline & $\alpha^{(0)}:=(0,0,0,0,0,0)^\top$ \\
		Feb 15 -- Mar 4 & Alerts & $\alpha^{(1)}:=(0.07,0.02,0.12,0.02,0.02,-0.01)^\top$ \\
		Mar 5 -- Mar 18 & School closures & $\alpha^{(2)}:=(0.055,0.09,0.15,-0.045,-0.015,0.01)^\top$ \\
		Mar 19 -- Jun 21 & School \& workplace closures & $\alpha^{(3)}:=(-0.3,-0.07,0.06,-0.42,-0.4,0.15)^\top$ \\
		\hline \hline
	\end{tabular}
	}
	\label{tab:medmob}
\end{table}

\begin{figure}[!ht]
	\caption{Simulated US COVID-19 cases with mobility controls $\bar{\alpha}_t\equiv\{\alpha^{(0)},\alpha^{(1)},\alpha^{(2)},\alpha^{(3)}\}$ (from top to bottom). The black curves are the median values and the shadow areas are bounded by the 0.45- and 0.55-quantiles.}
	\includegraphics[width=\linewidth]{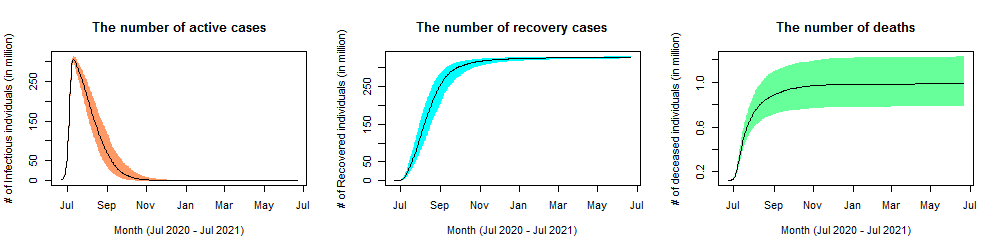} \\
	\includegraphics[width=\linewidth]{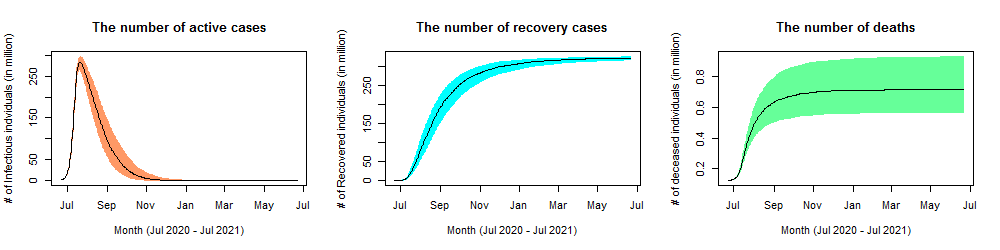} \\
	\includegraphics[width=\linewidth]{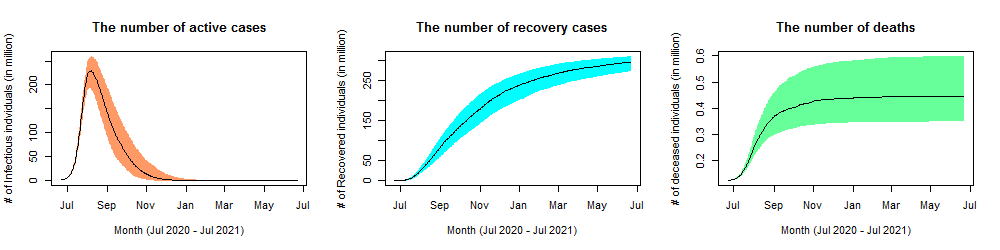} \\
	\includegraphics[width=\linewidth]{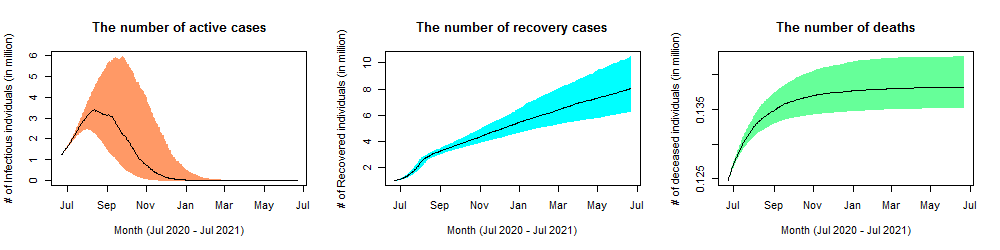}
	\label{fig:simIRD}
\end{figure}

Next, we simulate the SIRD model  \eqref{eq:sird}  and  \eqref{eq:ratesdynamics}  with parameters estimated on Jun 21 and the mobility indices listed in Table~\ref{tab:medmob}. Although the mobility indices are merely median values, we found that $\alpha^{(1)}$, $\alpha^{(2)}$, and $\alpha^{(3)}$ are close to the mobility indices of Feb 27, Mar 10, and May 6, respectively, in terms of Euclidean distance. For each community mobility, we conduct 10,000 simulations and compute the median curves of $I$, $R$, and $D$ over the subsequent one year (Jun 22, 2020 -- Jun 21, 2021). The 0.45- and 0.55-quantile curves are also included to visualize the volatility of the curves. The results are presented in  Figure~\ref{fig:simIRD} with different mobility controls.

From  Figure~\ref{fig:simIRD}, we can see that with the mobility chosen from Jan 3 to Mar 18, 2020, there would be many infectious cases and deaths. With the baseline mobility, more than 90\% of people in US will be infected.
With the responses in early March, the infectious cases and deaths are reduced but are still of a large magnitude. While the infectious cases and deaths are significantly reduced with the mobility of $\alpha^{(3)}$, the COVID-19 can be contained by the end of 2020.

We can also observe that the variances of $I$, $R$, and $D$ are time-inhomogeneous. Specifically, their variances are positively correlated with the current infectious cases $I$. It can be explained by noting the specification of the SIRD model \eqref{eq:sird}. Hence, controlling the mobility not only lowers the infection rate but also stabilizes the coronavirus situation.


\section{Deep-learning solution}\label{deep}

Such a complicated multi-objective stochastic optimal control problem in \eqref{eq:prob} does not admit an explicit analytical solution. It should be solved numerically and a promising computational approach is deep learning through an appropriate deep neutral network (DNN). As the DNN is closely related to the Markov decision process (MDP).
We transform the stochastic control problem \eqref{eq:prob} into a discrete-time MDP considered in \cite{Tsang2020}. More specifically,  we write the problem as follows
 \begin{equation}\label{eq:MDPprob}
  \begin{aligned}
  V_t(w) = & \min_{\alpha_s \in \mathcal{A}_s,~s=t,\ldots,t+h-1} J(t,X_t;\{\alpha_{s}\}_{s=t}^{t+h-1}),\\
  \hbox{s.t.}& \qquad w_{s} = \phi (w_{s-1} ,\alpha_{s-1}(w_{s-1}),\eta_{s}),~ s=t+1, \cdots,t+h,~w_t = w,
  \end{aligned}
 \end{equation}
where $w_s:= ( X_s, \widetilde{\alpha}_s   ) \in \mathbb{R}^{30}$ is the state variable; $\widetilde{\alpha}_s:= ( \alpha_i    )_{ i = s-4}^{s-1} \in \mathbb{R}^{24}$ is a vector of mobility indices over time-steps from $s-4$ to $s-1$;  
$\phi$ denotes the transition function given by the model \eqref{eq:sird} and \eqref{revise}, while $\widetilde{\alpha}_{s+1}$ is updated from $\widetilde{\alpha}_s$ through deleting $\alpha_{s-4}$ and adding $\alpha_s$; $\eta_{s}: =(Z_{s}^{\beta}, Z_{s}^{\gamma}, Z_{s}^{\delta})  \in \mathbb{R}^3$ is a vector of noisy information at time  $s$. The model is fully characterized by the state variable $w_s$ and thus the optimal control $\alpha_{s}$ depends only on the current state $w_s$. To simplify notation, we define the cumulative cost $C^{\lambda}_{\tau} = \sum_{s = t+1}^{\tau} \overline{c}^{\lambda}(s,X_s, \lambda;\alpha_{s-1}) $ for $t+1 \le \tau \le t+h$, where $\overline{c}^{\lambda}$ is defined in \eqref{revise}.
In the following, we detail the deep-learning solution to the MDP \eqref{eq:MDPprob}.  In fact, our deep-learning solution agrees with that proposed in  \cite{Tsang2020}, which establishes the theoretical convergence guarantee of their deep-learning solution and demonstrates the computation up to 100 dimensions. Our problem also encounters high-dimensionality but only of 30 dimensions, including the moving average controls. 
As mentioned before, the optimal control $\alpha$ is a function of current state $w$.  We represent this dependence by a multilayer feedforward neural network (FNN) with $L$ hidden layers in the following form:
\begin{align}\label{eq:dnn1}
w_s \in \mathbb{R}^{30} \rightarrow g_s^{L+1}(\theta_s)  \in \mathbb{R}^6 \rightarrow \alpha_s, \quad s= t,\cdots,t+h-1,
\end{align}
where $\theta_s$ denotes the parameters of the network. Let $\mathcal{D}^{L} = \{ g^{L+1} (\theta): \mathbb{R}^{30} \rightarrow \mathbb{R}^6  \}$ be the class of functions computed by the standard FNN.   $\mathcal{D}^{L} $ can be represented as follows:
\begin{align}\label{eq:dnn2}
g^0 &= w\cr
g^ {\ell}& = \sigma_{\ell} ( B^{\ell-1} g^{\ell-1} + M^{\ell-1}  ),~\ell = 1,\cdots,L,\cr
g^{L+1} &= B^{L} g^{L} + M^{L},
\end{align}
where $\sigma_{\ell}(\cdot)$ are non-linear activation functions, $ B^{\ell}$ is the matrix weight and  $M^{\ell} $ is the vector bias $\theta=(B^{\ell}, M^{\ell})_{\ell = 0}^{L}$. In this FNN, $g^0 \in \mathbb{R}^{30}$ is the input layer and  $g^{L+1} \in \mathbb{R}^6$ is the output layer, $\{g^{\ell}\}_{\ell = 1}^{L}$ are hidden layers for which the nonlinear activation functions are applied.  There are many possible choices of activation functions $\sigma_{\ell}(\cdot)$, such as the tanh activation function, rectified linear unit (ReLU), and the softmax activation function. 

To impose control bounds at the last step, we apply $\text{tanh}(x): = \frac{2}{1+e^{-2x}} -1 \in [-1,1]$ to the output layer $g^{L+1}$ element-wisely.  The class of the control functions computed by \eqref{eq:dnn1}  at time $s\in[t,t+h-1]$ is defined as
\begin{align}
\mathcal{G}_s= \{  \alpha_s:\mathbb{R}^{30} \rightarrow \mathbb{R}^6: \alpha_s (w_s;\theta_s) =  \text{tanh}( g^{L+1}_s ), ~ g^{L+1}_s \in \mathcal{D}^L \}.
\end{align}
As we consider feedback control policies $\alpha \in \mathcal{G}$, the relaxed problem becomes 
\begin{align}
V_t^*(w) &= \min_{\alpha_s \in \mathcal{G}_s,~s = t,\cdots,t+h-1} J(t,X_t;\{\alpha_s\}_{s=t}^{t+h-1}) \cr
&= \min_{\theta_s, ~s = t,\cdots,t+h-1} J(t,X_t;\{\alpha_s\}_{s=t}^{t+h-1}).
\end{align}
 We remark that the approximation of $V_t(w)$ by $V_t^*(w)$ is consistent with \cite{Tsang2020}.
Based on the above relaxed problem, the DNN architecture is illustrated as follows. 

\begin{algorithm}
		\caption{Deep neural network architecture}
		\begin{algorithmic}
	\REQUIRE ~~\\ 
	current state variable $w_t$, sampled random vectors $\{\eta_s\}_{s=t+1}^{t+h}$ 
	\ENSURE ~~\\
	Total cost $C^\lambda_{t+h}$
	\FORALL{timestep $s \in [t, t+h-1]$}
     \STATE{Build FNN learning layers of the control: $w_s \rightarrow g_s^0 \rightarrow g_s^1 \rightarrow \cdots \rightarrow g_s^{L+1} \rightarrow \alpha_s:$  
     $g_s^0 \gets w_s$}
     \FORALL{$\ell = 1: L$} 
     \STATE{$g_s^{\ell}  \gets \sigma_{\ell} (  B_s^{\ell} g_s^{\ell-1} +M_s^{\ell-1} ) $}
     \ENDFOR
     \STATE{ $g_s^{L+1} \gets B_s^L g_s^L + M_s^L$} 
     \STATE{$\alpha_s \gets \hbox{tanh}( g_s^L )$}
 
    \STATE{ Update the state variable using transition function
     $(w_s,\alpha_s,\eta_{s+1}) \rightarrow w_{s+1}:  w_{s+1} = \phi(w_s,\alpha_s,\eta_{s+1}) $}
     \STATE{Compute cumulative cost function 
     $(w_{s+1},\alpha_s, C^\lambda_{s}) \rightarrow C^\lambda_{s+1}: C^\lambda_{s} + \overline{c}^{\lambda}(s+1, X_{s+1}, \lambda;\alpha_{s} ) $}
    \ENDFOR
    \end{algorithmic}
\end{algorithm}

We would like to stress that our empirical computation has PC constraints on $\alpha_s$ for enhancing the computational efficiency. In other words, we optimize the cumulative cost $\overline{C}^\lambda_t$ for which we perform cross-validation to identify the best fit penalty coefficients $\overline{\lambda}_i$ for $i=1,2, \cdots, 6$. With the architecture above, we sample  $\{\eta_s\}_{s=t+1}^{t+h}$ as the input data and train the DNN with standard stochastic gradient descent (SGD) method. The training algorithm can be easily implemented using common libraries (e.g., TensorFlow \cite{Tensorflow2015}).

We use Keras in TensorFlow to implement our algorithm with the Adam optimizer \cite{Kingma2015} to optimize parameters. For the architecture of the FNN, we set the number of hidden layers to 2, i.e, $L = 2$. We choose the ReLU as our activation function for the hidden layers.  We set the number of neurons for each hidden layer to 64, the initial learning rate to 0.0001. The model is trained with 20,000 simulated training samples for 100 times. After training, a set of simulated testing samples is applied to the model to obtain a set of optimized controls.
\end{appendix}

\end{document}